# Multispectral household plastic classification for recycling using a camera array

Katja Kossira[1*], Jürgen Seiler[1†] and André Kaup[1†]

[†]J¨urgen Seiler and Andr´e Kaup have equally contributed to this work.

*Correspondence: katja.kossira@fau.de

[1] Chair of Multimedia Communications and Signal Processing, Friedrich-Alexander University Erlangen-Nürnberg, Cauerstraße 7, 91058 Erlangen, Bavaria, Germany

**Abstract**
Plastic pollution has become a persistent problem in natural ecosystems. Driven by insufficient waste management, limited recycling efficiency, and high costs for recycling, plastic waste in the environment poses significant ecological and health risks. Recycling requires accurate identification of polymer types, but existing optical sorting systems often struggle to distinguish common household plastics. In this work, we present a novel classification approach based on a multispectral imaging system consisting of nine cameras equipped with near-infrared bandpass filters. The system is designed to discriminate the seven most common household plastics. From the resulting multispectral images, we extract the spectral fingerprints and derive features such as intensity differences between specific wavelength pairs and their slopes, as well as false-color image representations. A dedicated preprocessing pipeline aligns and normalizes the data before classification. We recorded a multispectral household plastic database (https://github.com/FAU-LMS/MHPM) and trained four different classifiers Gradient Boosting, Extreme Gradient Boosting, Light Gradient Boosting Machine, and CatBoost. The best-performing model achieves a classification accuracy of 86.7%. The computational runtime is 2.603 $\mu$s per pixel, enabling efficient processing of high-resolution images. The entire setup is built from off-the-shelf hardware components, which makes replication straightforward and allows direct integration into industrial sorting pipelines.

**Keywords:** Multispectral imaging, Camera arrays, Plastic classification, Recycling

## 1 Introduction

Over the past decades, global plastic production has been steadily increasing from 2 million tons in 1950 to 391 million tons in 2021 [1]. This development corresponds to a compound annual growth rate (CAGR) of 7.7%, meaning that global production has nearly doubled every 10 years. Overall, today's output is about 195 times higher than in 1950. With plastic packaging being the largest market sector of plastic resins, these products are designed for immediate disposal [2]. Consequently, plastic pollution has emerged as one of the most pressing environmental challenges of our time, severely impacting ecosystems, wildlife, and human health. In addition to its environmental persistence, plastic production itself contributes significantly to global greenhouse gas emissions due to its reliance on fossil fuel-based manufacturing processes [3]. In 2020,

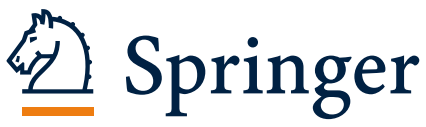

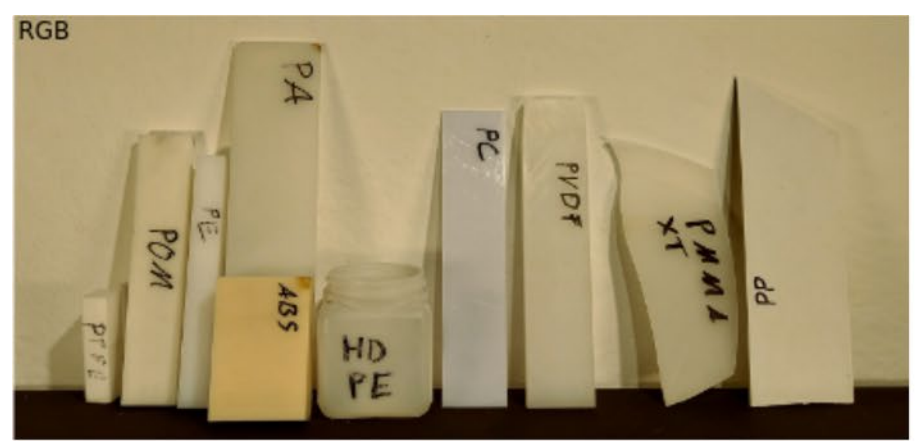

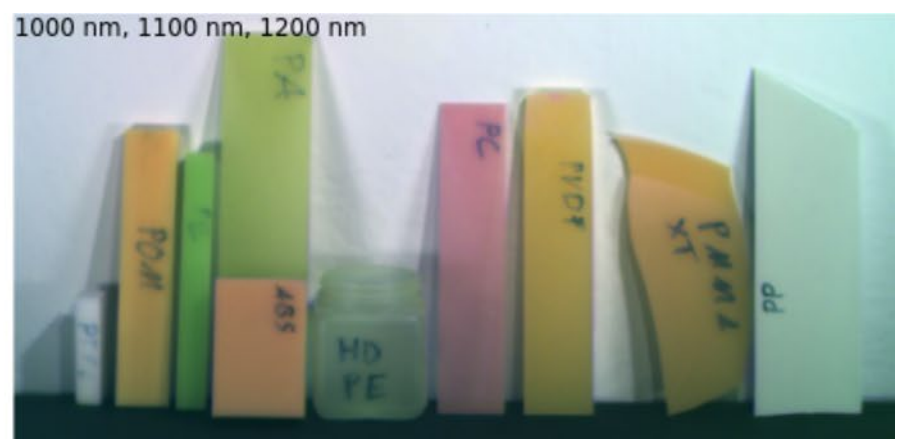

**Fig. 1** Original RGB image (left) and false-color image after superimposition of the three infrared channels 1000 nm, 1100 nm, and 1200 nm as Red, Green, and Blue channel (right). Different plastics appear in different colors in the IR and, hence, can be distinguished

the processes of plastic production, conversion, and waste management were responsible for approximately 2450 million tons $CO_2$-equivalent emissions, accounting for 5% of the world's industrial greenhouse gas output [4]. At the same time, millions of tons of plastic waste enter the environment each year. Hence, a growing global awareness of the urgent need for effective and sustainable waste management solutions arises.

In response to the increasing volumes of plastic waste, three main management strategies are applied. These methods include landfilling, incineration, and recycling. In 2020, they accounted for 39%, 24%, and 22% of global plastic waste, while 15% was mismanaged [3]. Landfilling and incineration reduce waste volumes but cause long-term environmental burdens such as groundwater contamination [5], greenhouse gas emissions [6], and hazardous residues [2]. In contrast, recycling avoids these drawbacks by reusing existing material, thereby reducing dependence on virgin fossil resources and cutting greenhouse gas emissions by about 700 million tons annually [7]. Given these benefits, recycling must be considered the central strategy for mitigating plastic pollution.

However, despite its environmental and ecological advantages, recycling systems still face considerable practical challenges. Effective recycling depends on the accurate identification and separation of materials, which is often compromised by contamination, damage, dirt, or obscured labeling on plastic products. While many plastics are marked with standardized identification codes, these are often not visible, readable, or recognizable at all. As a result, even advanced sorting systems struggle with reliable classification.

To overcome these identification bottlenecks, we introduce a multispectral imaging approach. A camera array, each camera equipped with a distinct near-infrared (NIR) bandpass filter, records the unique spectral signatures of polymer carbon chains, which absorb and reflect NIR wavelengths in material-specific patterns. By capturing these intrinsic spectral fingerprints across multiple bandwidths, subtle differences between plastic materials can be visualized. Figure 1 provides an example of how different polymers appear in distinct colors when selected NIR channels are combined into the Red, Green, and Blue channels of a false-color image.

We present a system concept for multispectral household plastic classification (MHPC). The central contributions are the introduction of a new multispectral household plastic database,[1] a multispectral camera array with near-infrared bandpass filters, a preprocessing pipeline, the system integration of the hardware and software components into a functional setup, and a classification model for the multispectral household plastic classification. The dataset contains multispectral recordings of the seven most common household plastic types which were used to train and validate the classifiers. The preprocessing pipeline aligns and normalizes the raw images and extracts fingerprints and derived features such as intensity differences between wavelengths and false-color image representations. The MHPC model maps these features to the respective plastic types. By exploiting both spatial and spectral information, the pipeline improves discrimination between materials. Unlike methods that rely on pre-printed resin identification codes [8] or manual pre-cleaning, this classification is based on the inherent optical properties of the polymers. The recorded database served exclusively for training, and classifier performance was assessed both on the dataset and on separate plastic objects. The proposed setup demonstrates that multispectral image signals can be effectively used for the classification of plastics and provides a basis for their application in industrial sorting systems and recycling.

[1] https://github.com/FAU-LMS/MHPM

## 2 Waste management solutions for recycling

Accurate polymer identification and separation is crucial for recycling, as the diverse carbon backbone structures of different polymers require material purity for reprocessing. Various technologies are currently in use to address this issue [5].

### 2.1 Common identification and sorting methods

Many plastic products are labeled with Resin Identification Codes (RICs) [8], which are standardized numeric symbols (usually 1–7) placed within a triangle of arrows and embossed directly on the plastic surface. These codes indicate the polymer type (1 = PET, 2 = HDPE, 3 = PVC, 4 = LDPE, 5 = PP, 6 = PS, and 7 = other) and are intended to aid in manual or semi-automated sorting. However, manual sorting is slow and error-prone, and codes may be damaged, missing, or obscured. They also lack information on additives or composites. Barcodes and QR codes can provide more data and are machine-readable [9], yet require clean, intact surfaces and are not universally standardized.

Density-based separation exploits buoyancy differences [10], as PP and PE float while PET sinks. This approach is suitable as a secondary sorting step after shredding but is limited to materials with distinct densities. Further, wet material is produced that requires drying and additional expensive machinery [11], which increases time and costs.

### 2.2 Advanced optical and spectral technologies

While techniques like multispectral photoacoustic imaging [12] use light-induced sound waves to test material properties, a more widely adopted and more advanced technique is near-infrared (NIR) spectroscopy. It identifies materials based on their characteristic reflection spectra under NIR illumination [13]. However, reliable identification requires multiple point measurements per object, especially in cases of heterogeneous surfaces or composite items such as PET bottles with HDPE caps. This increases the effective measurement time and can limit throughput in high-volume scenarios. Additionally, setup and calibration demand precise spectral libraries and can be cost-intensive [14]. In practice, NIR spectroscopy has mainly been applied to distinguish a limited set of polymers, often not more than five to ten different plastic types.

To overcome these limitations, multispectral and hyperspectral imaging (MSI/HSI) cameras have been introduced into industrial applications [15] as well as environmental approaches for plastic detection [16]. In parallel, push-broom sensors [17, 18] have been employed in plastic sorting environments. These imaging systems combine spatial and spectral information, allowing for the detection of localized features such as contamination or composite structures. By capturing entire scenes, they provide a more comprehensive material profile and enable more precise sorting. A further advantage of MSI is that only a limited number of spectral bands are acquired. These bands are usually selected to be decorrelated [19], i.e., they provide complementary information with minimal redundancy, thereby reducing data volume and enabling faster processing. However, as these imaging systems often rely on specialized, non-standardized hardware, their integration into widespread industrial sorting processes remains limited and expensive. Furthermore, despite their advanced capabilities, classification accuracy tends to decrease as the number of plastic types increases.

Other current methods have explored alternative approaches that combine cost-efficient multispectral NIR sensing with machine learning. West et al. [20] present a low-cost multispectral NIR sensor system which, in combination with supervised learning algorithms, achieved a classification accuracy of up to 62.08% for several common polymers. The reduced accuracy compared to full spectroscopic systems such as [13] or [14] is mainly due to the limited number of spectral bands, yet the system is attractive because of its low cost and ease of deployment. Werner et al. [21] further optimized this idea by applying a band selection strategy in the NIR spectrum together with a k-nearest neighbor classifier. They reduced hardware complexity and data volume while still achieving an average classification accuracy of 90% across five different plastic types. This demonstrates that targeted feature extraction and algorithmic optimization can partly offset the limitations of low-cost sensors. In summary, these studies indicate that while low-cost multispectral systems inherently perform worse due to fewer available bands, they offer significant advantages in terms of affordability and scalability. At the same time, their reliance on a small set of bands makes them sensitive to spectral variations introduced by additives or surface contamination, and the restriction to a maximum of five polymer types limits their applicability in realistic sorting scenarios.

## 3 Data acquisition

To overcome the limitations of existing systems, we developed a custom multispectral imaging setup capable of reliably classifying the seven most common plastics in household waste, rather than the five typically handled by conventional systems. The hardware is designed for flexibility and scalability, and off-the-shelf components make it affordable for a wide range of applications. As the setup does not use a single multispectral camera but a camera array with multiple filters in front of each camera, a data preprocessing pipeline is essential to ensure the spatial alignment of the individual images.

### 3.1 Hardware description

The hardware setup consists of two halogen lamps for infrared illumination and a camera array equipped with infrared bandpass filters mounted in front of the lenses. A side view of the recording setup is shown in Fig. 2 a. The two 400 W halogen lamps[2] are placed on either side of the system and provide a broad and continuous spectrum ranging from 300 nm to 2000 nm. Their position and angle can be adjusted individually to optimize illumination conditions and minimize specular reflections on the plastic samples. As halogen lamps produce significant heat, they are positioned laterally outside the immediate camera enclosure rather than directly next to the camera array. This placement ensures adequate air circulation and protects the cameras from overheating and faster hardware degradation. Further, this controlled environment is not merely a laboratory constraint, but strictly represents the required operational conditions for an industrial setup, as the infrared imaging modality does not permit ambient lighting.

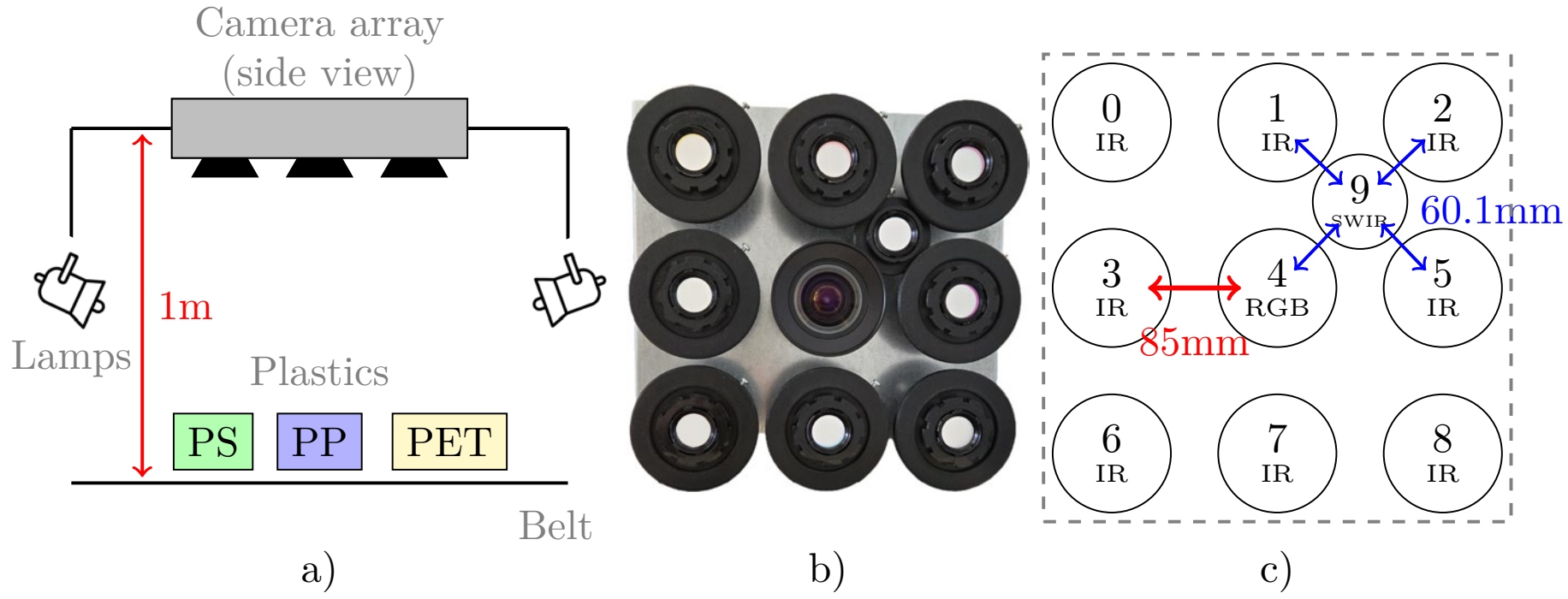

**Fig. 2** **a** Sketch of hardware setup used for capturing the multispectral household plastic images, **b** camera array, and **c** arrangement of cameras within the array. IR: grayscale camera with infrared filter, RGB: RGB camera, and SWIR: short-wave infrared camera with infrared filter

To ensure consistent brightness and account for lighting fluctuations, three Zenith Polymer calibration standards[3] are included in each captured frame. These reference targets provide defined reflectance values of 0%, 50%, and almost 100% across a spectral range of 250 nm–2450 nm for black, gray, and white reference targets, respectively.

The camera array is mounted above a fixed capture area at a distance of 1 m to the plastic samples. Its design follows the principles of the camera array for multispectral imaging (CAMSI) [22]. It consists of eight peripheral monochrome (grayscale) cameras, one RGB camera in the center, and one short-wave infrared (SWIR) camera. The array itself is shown in Fig. 2 b, and the camera arrangement is detailed in Fig. 2 c. Cameras 0–8 are placed in a 3×3 grid with horizontal and vertical baselines of 85 mm. Camera 9 is mounted diagonally above the central camera with a baseline of 60.1 mm. The grayscale cameras are sensitive to wavelengths between 300 nm and 1150 nm. The RGB camera 4 is used to obtain a true-color image of the scene. Camera 9 extends the spectral range with a sensitivity between 400 nm and 1700 nm. Detailed hardware specifications are provided in Table 1.

Infrared bandpass (BP) filters are mounted in front of the grayscale and SWIR camera lenses. These filters are easily interchangeable and feature a bandwidth of 50 nm, ±25 nm around the central wavelength, with an optical density of 4. The selected filters for the grayscale cameras correspond to the central wavelengths 800 nm, 850 nm, 900 nm, 950 nm, 975 nm, 1000 nm, 1050 nm, and 1125 nm. For the SWIR camera, a 1350 nm filter is used. The filter selection was performed using the conditional filter band selection (CFBS) [23] and is explained in more detail in Sect. 4.2.

[2] OSRAM Haloline 400 W.

[3] SphereOptics Zenith Polymer.

**Table 1** Camera parameters of recording setup

| System parameters | Camera array | | |
|---|---|---|---|
| | **Grayscale** | **RGB** | **SWIR** |
| Camera type | Allied Vision Alvium 1800 U-1240 m | Allied Vision Alvium 1800 U-1240c | Allied Vision Alvium 1800 U-130 m VSWIR |
| Baseline B | 85 mm | 85 mm | 60.1 mm |
| Focal length | 5 mm | 5 mm | 5 mm |
| Image sensor | Sony IMX226 | Sony IMX226 | Sony IMX990 |
| Sensor pixel size | 1.85$\mu$m | 1.85$\mu$m | 5$\mu$m |
| Resolution | 4024 x 3036 | 4024 x 3036 | 1296 x 1032 |
| Frame rate | 35 fps | 35 fps | 130 fps |
| Bit depth | 10 | 10 | 10 |
| Lens transmission | 300 nm–1150 nm | 300 nm–1100 nm | 400 nm–1700 nm |

Our modular hardware setup offers several practical advantages. Unlike existing systems such as [13, 16] or [17] that rely on specialized hardware, our setup is built using off-the-shelf components. This significantly simplifies replication, making the system accessible for a wide range of research and industrial applications. Moreover, the modular design allows a flexible arrangement of components and easy scaling, as cameras, and thus infrared wavelengths, can be added or removed depending on the classification requirements. In case of hardware failure, individual components can be replaced independently, which reduces maintenance costs. While the setup is physically larger than a single multispectral camera or spectrometer, it is still compact relative to the scale of industrial sorting facilities.

### 3.2 Data preprocessing

The described hardware setup captures images at different wavelengths simultaneously. Due to the physical arrangement of the array, the individual cameras exhibit slight spatial offsets, resulting in horizontal, vertical, and diagonal misalignments between the wavelength channels. Consequently, several preprocessing steps are required to spatially align and combine all channels into a single multispectral image that accurately combines all spectral components for subsequent feature extraction and classification. Further preprocessing includes color sensitivity correction and intensity correction. The color sensitivity correction compensates for variations in sensor responses between individual cameras, which may occur due to sensor noise, manufacturing tolerances, or device aging, while the intensity normalization adjusts for differences in exposure time across wavelengths. The complete preprocessing pipeline is illustrated in Fig. 3, and each step is explained in detail below.

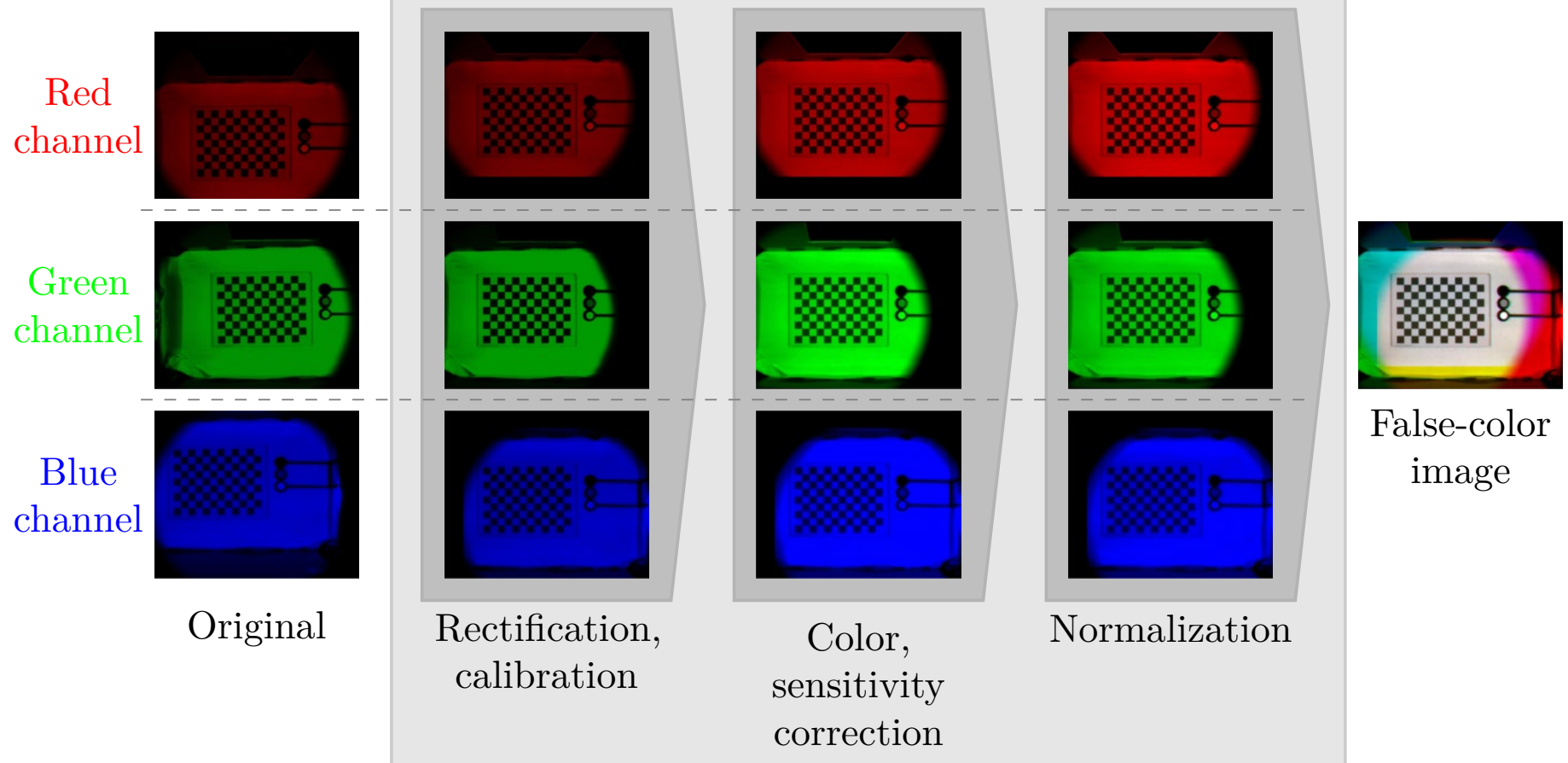

**Fig. 3** Preprocessing pipeline. For the example, three infrared images captured with camera 1, 5, and 6 are mapped to the R, G, and B components of a false-color image. The principle applies for all infrared images

#### *3.2.1 Rectification and calibration*

Due to the physical displacement of the cameras, spatial misalignment between the captured images occurs. First, it is not possible to perfectly align the cameras onto the hardware mount. Second, the image sensors might not be installed precisely enough in the camera body. Hence, the same physical point projects to different pixel locations across the set of images. Consequently, to calculate a multispectral image, all cameras must be calibrated with respect to the center camera first, assuming that intrinsic camera parameters and the focus are already compensated using methods like [24, 25] and [26]. Corresponding pixels are misplaced both in horizontal and vertical directions in the recorded set of images. By recording a calibration pattern in front of all cameras, it is possible to estimate the homography of the planes. Thus, the displacement for all cameras can be corrected, and the peripheral cameras are warped to the center. A checkerboard was used as rectification and calibration pattern, as this is a well-investigated approach. The derivation of the calibration matrices and the transformation is performed analogously to [22].

Since this process corrects the images for the specific calibration depth, objects closer or farther than the calibration plane will not align perfectly after rectification and calibration. However, these deviations decrease as the distance from the camera increases, because the angular disparity between views reduces with depth. In our setup, the plastic objects are relatively flat and placed at a sufficient distance from the multispectral camera array. Consequently, these residual misalignments are minimized and thus negligible.

While this simplified approach is effective for flat items, larger 3D objects such as bottles or containers introduce physical parallax errors at object boundaries due to their size. To evaluate the impact of these boundary misalignments, advanced 3D registration and reconstruction algorithms, specifically Cross Spectral Reconstruction (CSR) [22], Multispectral Image Registration (MSIR) [27], and Deep Guided Neural Network (DGNet) [28], were integrated in the setup. These methods significantly increased computational time without yielding a measurable improvement in the final classification accuracy. Although misaligned pixels at the object borders can occasionally produce corrupted pixel-wise spectral signatures, these local edge artifacts represent only a minor fraction of the object's total surface. Because the classification pipeline is designed to derive the final object-level prediction through a general majority vote over the object's area, such minor pixel misclassifications do not impact the overall result. For this reason, rectification and calibration suffice, and full reconstruction as in [22, 27] or [29] are not required.

#### 3.2.2 Color and sensitivity correction

To evaluate spectral responses of different plastics reliably, consistent sensitivities across the employed cameras' sensors are indispensable. Due to device altering, thermal issues or electrical noise, inconsistencies can occur, even though the cameras are of the same make and model. To compensate for that, inter-camera sensitivity correction with a median consensus image [30] is applied. The inclusion of all available images in the reference leads to more realistic results. The correction process has to be performed once after the camera array is started.

#### 3.2.3 Normalization

To ensure comparability between the spectral channels and to compensate for exposure differences between the individual cameras and their filters, normalization is essential. Even with careful tuning, slight variations in sensor sensitivity, exposure settings, or filter transmission can lead to inconsistent intensity values across bands. To normalize the images, a Min-Max Normalization [31] as in the following equation is applied:

$$I_{\text{normalized}} = \frac{I - \min_{\text{int}}}{\max_{\text{int}} - \min_{\text{int}}} \quad . \tag{1}$$

Here $I$ is the raw captured image, and $\min_{\text{int}}$ and $\max_{\text{int}}$ are the minimum and maximum intensities in the image, respectively. They are obtained from the Zenith Polymer targets as follows:

$$\begin{aligned} \max_{\text{int}} &= \text{white Zenith-Polymer}, \\ \frac{\max_{\text{int}}}{2} &= \text{gray Zenith-Polymer}, \\ \min_{\text{int}} &= \text{black Zenith-Polymer} \end{aligned} \tag{2}$$

Through dedicated calibration measurements using the ISO-certified Zenith Polymer patches, we confirmed that the gray standard reflects exactly half the intensity of the white standard, which means that the white-to-gray ratio is precisely 2. Because the white patch can saturate and clip at the sensor's maximum value, directly measuring $\max_{\text{int}}$ is not optimal. Instead, we use the white-to-gray ratio of (2) to reformulate (1):

$$I_{\text{normalized}} = \frac{I - \min_{\text{int}}}{\frac{\max_{\text{int}}}{2} - \min_{\text{int}}} \cdot 128 \quad . \tag{3}$$

### 3.3 Dataset

Using this setup and the described preprocessing steps, we recorded and provide a multispectral household plastic (MHP) database, which was subsequently used to train the classifiers. The ground truth plastic type was determined by manually checking the resin identification codes embossed on each item. For composite items, the material corresponding to the dominant component was chosen as the class label. For instance, in beverage bottles, the PET body was considered the representative material over HDPE caps or labels. The MHP database comprises 11 different types of plastics commonly found in household waste, as summarized in Table 2. For each individual plastic type, between four and twelve plastic samples have been recorded.

The prefix “R” refers to “recycled,” indicating that the material has undergone at least one recycling process but retains the same chemical composition as its virgin counterpart. The suffix “G” in PET-G refers to glycol, which is added to increase its flexibility. Measurements have shown that recycled and virgin materials, as well as plastics with additives, exhibit nearly identical spectral characteristics due to their unchanged polymer structure. Hence, we treat them as a single class with slightly increased spectral variance and group them accordingly, resulting in seven final classes containing between 6 and 19 samples per class.

To ensure data quality, all plastic items were thoroughly cleaned prior to imaging. Residual contamination is known to affect spectral measurements primarily in transparent containers, where residues can interfere with light transmission and distort the recorded spectra. In contrast, for opaque plastics, such contamination has little to no impact on the measured reflectance. Since all objects in the dataset were cleaned, contamination effects are considered negligible. Moreover, differences in object shape were found to have minimal influence on the spectral signature and are thus not considered a relevant factor in classification.

Each plastic class in the MHP database includes several distinct objects. Representative examples comprise empty and cleaned shampoo or water bottles, yogurt cups, chewing gum containers, or chocolate packaging. Selected samples are shown in Fig. 4, illustrating the material appearance at 14 selected wavelengths. These 14 bands were chosen from the full set of 24 recorded wavelengths to visually demonstrate the characteristic spectral appearance of the materials.

**Table 2** List of dataset plastic types, abbreviations, and individual and grouped sample counts

| Abbreviation | Chemical name | #samples | #grouped |
|---|---|---|---|
| HDPE | High-density polyethylene | 8 | } 12 |
| R-HDPE | Recycled high-density polyethylene | 4 | |
| LDPE | Low-density polyethylene | 4 | } 8 |
| R-LDPE | Recycled low-density polyethylene | 4 | |
| PET | Polyethylene terephthalate | 7 | 3 } 19 |
| PET-G | Polyethylene terephthalate glycol-modified | 6 | |
| R-PET | Recycled polyethylene terephthalate | 6 | |
| PP | Polypropylene | 12 | 12 |
| PS | Polystyrene | 6 | 6 |
| PTFE | Polytetrafluoroethylene | 7 | 7 |
| PVC | Polyvinyl chloride | 6 | 6 |

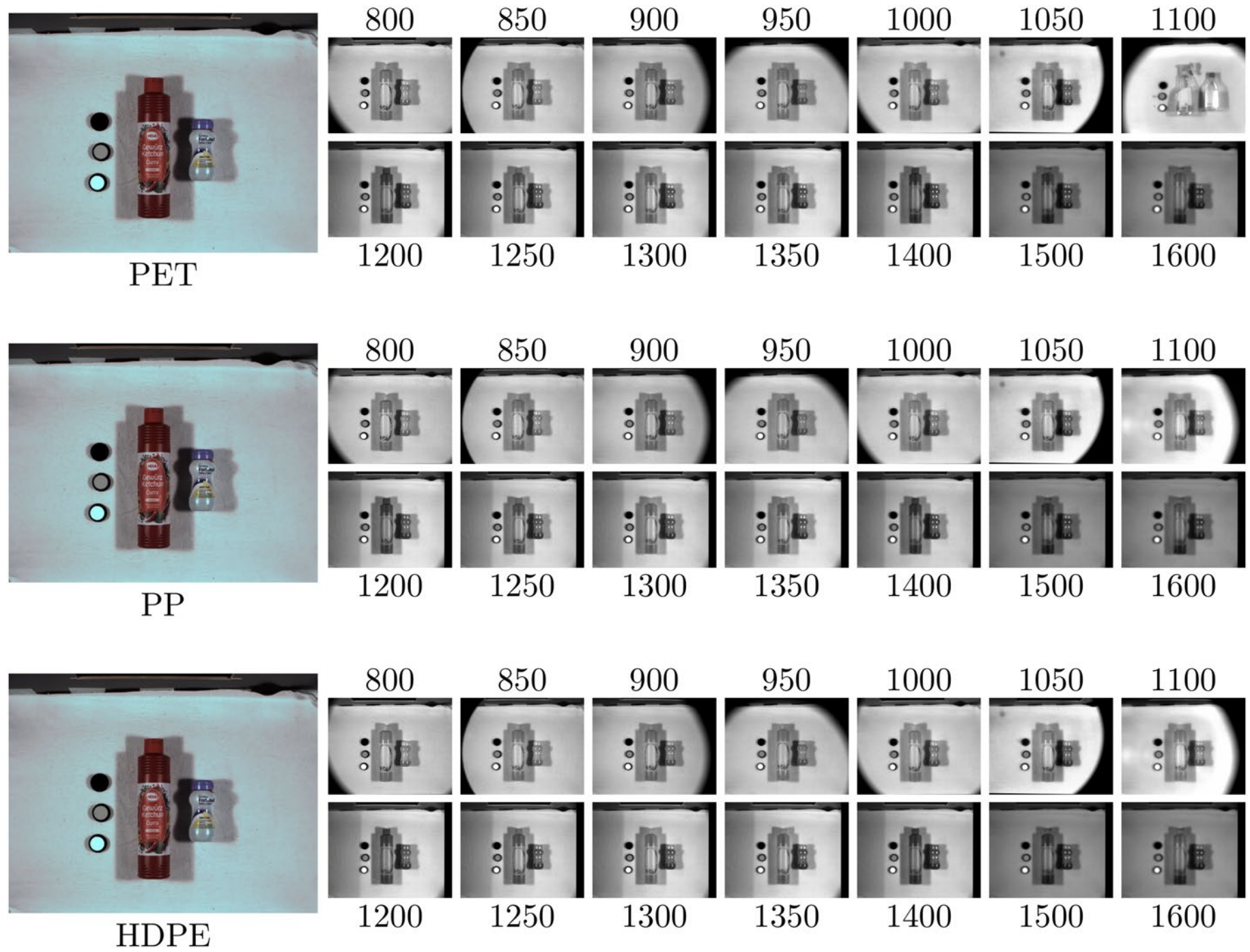

**Fig. 4** Representative examples of household plastics included in the MHP database, shown at 14 selected wavelengths, and the original RGB image on the left. Each sample was recorded at 24 wavelengths in total, covering the spectral range from 800 nm to 1600 nm

For every object, we recorded an RGB image and multispectral images at all 24 wavelengths, as listed in Table 3. The multispectral acquisition was performed by manually exchanging the optical filters between all captures until all wavelengths were recorded. Recording the complete spectral near-infrared range enables a flexible, application-dependent wavelength analysis. However, in practical implementations, only a small amount of filters can be used simultaneously due to the hardware constraints of the multispectral camera array. Consequently, in this work, the nine most informative wavelengths were later selected through an optimization process described in the subsequent Sect. 4.2.

In addition to variations in object shape and packaging design, the dataset also includes differently colored plastic samples in each class. Tests were conducted to assess the impact of these visual and spectral differences. While increased intra-class variance due to color and surface texture was observed, the underlying spectral signatures remained sufficiently distinctive to enable meaningful classification. The recorded spectral intensities across the different wavelengths ultimately serve as input for the feature extraction and classification models described in the following Sect. 4.

## 4 Methodology

### 4.1 Feature extraction

Multispectral imaging provides a non-destructive and efficient approach for the classification of plastics based on their spectral properties. To facilitate accurate classification, various features can be extracted from the measured data, such as color histograms to describe the overall material distribution, texture descriptors, or morphological shape contours. However, as our system is based on a pixel-wise classification followed by a majority voting for real-time sorting, such global features are computationally expensive [23] and less effective for high-throughput scenarios [14]. Therefore, we focus on spectral and local intensity features derived from the nine selected filter wavelengths $\Lambda = \{\lambda_1, ..., \lambda_9\}$. The extracted features are concatenated into a single joint 1D feature vector $\mathbf{x} \in \mathbb{R}^{277}$ for each pixel, and the exact construction is detailed below. Spectral fingerprints, slopes, and false-color images extracted from the MHP dataset serve as reference features for training the classifiers.

**Table 3** Filters used for recording the dataset

| Description | Bandwidth (nm) | Optical density | Center frequency (nm) | Transmission (nm) |
|---|---|---|---|---|
| Steep bandpass | 50 | 4 | 800 | 775–825 |
| Steep bandpass | 50 | 4 | 825 | 800–850 |
| Steep bandpass | 50 | 4 | 850 | 825–875 |
| Steep bandpass | 50 | 4 | 875 | 850–900 |
| Steep bandpass | 50 | 4 | 900 | 875–925 |
| Steep bandpass | 50 | 4 | 925 | 900–950 |
| Steep bandpass | 50 | 4 | 950 | 925–975 |
| Steep bandpass | 50 | 4 | 975 | 950–1000 |
| Steep bandpass | 50 | 4 | 1000 | 975–1025 |
| Steep bandpass | 50 | 4 | 1025 | 1000–1050 |
| Steep bandpass | 50 | 4 | 1050 | 1025–1075 |
| Steep bandpass | 50 | 4 | 1075 | 1050–1100 |
| Steep bandpass | 50 | 4 | 1100 | 1075–1125 |
| Steep bandpass | 50 | 4 | 1125 | 1100–1150 |
| Steep bandpass | 50 | 4 | 1150 | 1125–1175 |
| Steep bandpass | 50 | 4 | 1200 | 1175–1225 |
| Steep bandpass | 50 | 4 | 1250 | 1225–1275 |
| Steep bandpass | 50 | 4 | 1300 | 1275–1325 |
| Steep bandpass | 50 | 4 | 1350 | 1325–1375 |
| Steep bandpass | 50 | 4 | 1400 | 1375–1425 |
| Steep bandpass | 50 | 4 | 1450 | 1425–1475 |
| Steep bandpass | 50 | 4 | 1500 | 1475–1525 |
| Steep bandpass | 50 | 4 | 1550 | 1525–1575 |
| Steep bandpass | 50 | 4 | 1600 | 1575–1625 |

#### *4.1.1 Spectral fingerprints*

First, the spectral signatures, also called spectral fingerprints, shown in Fig. 5, are calculated, as plastics exhibit material-specific reflectance characteristics across different wavelengths. These signatures serve as primary descriptors for differentiating plastic types.

To obtain the ground truth spectral fingerprints, the MHP dataset images are processed for each plastic material and wavelength. For every wavelength band, the corresponding reflectance values are extracted from the image regions assigned to the respective plastic. To reduce sensor noise and local inhomogeneities, a sliding window approach with 7×7

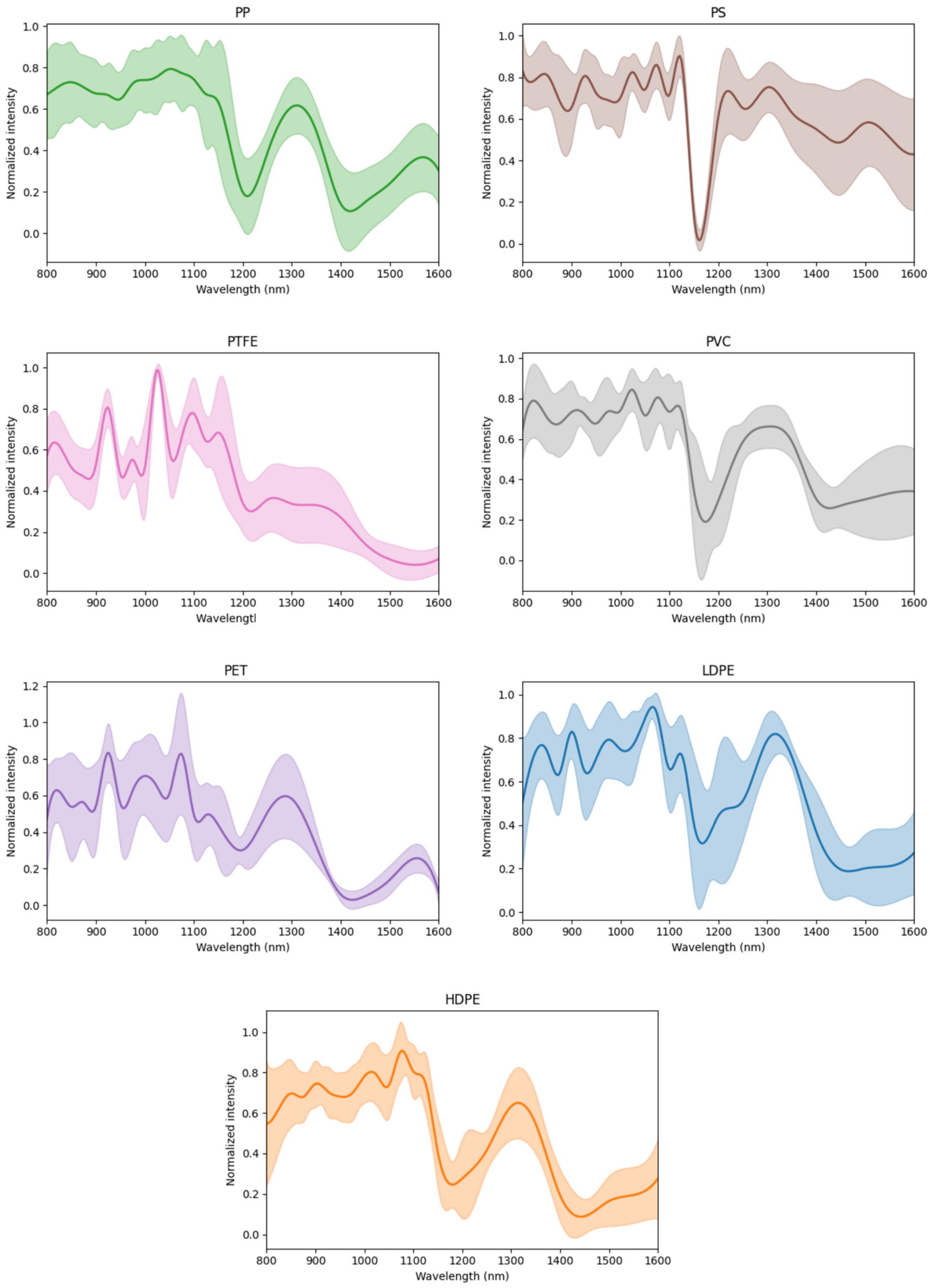

**Fig. 5** Spectral fingerprints of tested plastics. The solid line represents the median, while the transparent shaded area around it indicates the standard deviation

pixels is applied. Several different window combinations and sizes were tested, and the selected window size produced the best results.

The resulting spectra are interpolated using cubic spline interpolation to achieve a uniform spectral resolution across all materials. Subsequently, each spectrum is normalized to the [0,1] range according to

$$R_{\text{norm}}(\lambda) = \frac{R(\lambda) - R_{\min}}{R_{\max} - R_{\min}} \quad , \tag{4}$$

where $R(\lambda)$ is the measured reflectance at wavelength $\lambda$.

For each material, multiple normalized spectra are obtained from individual measurements. From these, the mean spectral signature $\bar{R}(\lambda)$ and the standard deviation $\sigma(\lambda)$ are calculated as

$$\begin{aligned} \bar{R} &= \frac{1}{N}\sum_{i=1}^{N} R_i(\lambda) \ , \\ \sigma(\lambda) &= \sqrt{\frac{1}{N-1}\sum_{i=1}^{N}(R_i(\lambda) - \bar{R}(\lambda))^2} \ , \end{aligned} \tag{5}$$

where $N$ denotes the number of individual spectra. The mean curves represent the characteristic reflectance behavior of the material, while the standard deviation $\sigma$ quantifies the variability between different samples, e.g., due to color additives or production differences. Figure 5 shows the spectral fingerprints of all tested plastics, visualized as mean reflectance curves with shaded areas indicating $\pm\sigma$ confidence intervals. As stated earlier in Sect. 3.3, the recycled plastics are grouped along with their virgin counterparts. It is evident that the different materials can be distinguished based on their spectral fingerprints, especially in the near-infrared range between 800 nm and 1350 nm. For example, PP consistently exhibits low reflectance around 1200 nm, which increases at higher wavelengths before dropping again. Thus, the primary component of the feature vector is the intensity vector $\mathbf{R} \in \mathbb{R}^9$:

$$\mathbf{R} = [R(\lambda_1), R(\lambda_2), ..., R(\lambda_9)]^{\mathsf{T}} \tag{6}$$

While this vector directly captures the absolute pixel-wise reflectance values, additional localized descriptors are necessary to model structural variations between adjacent bands. Hence, the spectral slopes $\Delta$ between adjacent bands are computed as well to capture the dynamics of reflectance changes. For example, LDPE shows a steep increase in reflectance between 800 nm and 850 nm, while PP exhibits a flatter progression. These slopes serve as additional discriminative features, complementing the information from the absolute reflectance values. Furthermore, to ensure robustness against multiplicative intensity fluctuations, nonlinear spectral intensity ratios $\mathbf{Q}$ are introduced. Formally, the vector of spectral slopes $\Delta \in \mathbb{R}^8$ and the complementary vector of the spectral intensity ratios $\mathbf{Q} \in \mathbb{R}^8$ between adjacent bands are defined as follows:

$$\Delta = \left[\frac{R(\lambda_2) - R(\lambda_1)}{\lambda_2 - \lambda_1}, \ldots, \frac{R(\lambda_9) - R(\lambda_8)}{\lambda_9 - \lambda_8}\right]^{\mathsf{T}}, \quad \mathbf{Q} = \left[\frac{R(\lambda_2)}{R(\lambda_1)}, \ldots, \frac{R(\lambda_9)}{R(\lambda_8)}\right]^{\mathsf{T}} \ , \tag{7}$$

where a safe division is utilized if the denominator $R(\lambda_i) = 0$.

#### *4.1.2 False-color images*

To enhance material discrimination, additional features are extracted from false-color representations of the dataset images.

False-color images are generated by mapping three selected infrared channels to the red, green, and blue channels, respectively. Plastics appear in distinct colors depending on the specific combination of wavelengths used for the RGB overlay. Figure 6 illustrates examples of false-color images generated using different infrared channels. While global spatial statistics such as color histograms can be derived from these images and served as a valuable comparative baseline during our model development, they do not flow into the final feature vector to maintain the computational efficiency of the pixel-wise classification. From each resulting false-color image, the representative color of the plastic material is extracted and used as a feature for the classifier. Thus, all possible three-channel combinations from the available wavelengths are considered to maximize the separability between plastic types. For the nine available wavelengths, there are exactly $\binom{9}{3} = 84$ unique triplets. For each triplet combination $k = (\lambda_r, \lambda_g, \lambda_b)$, a three-dimensional color intensity subvector $\mathbf{f}_k$ is defined based on the reflectance values

$$\mathbf{f}_k = [R(\lambda_r), R(\lambda_g), R(\lambda_b)]^\mathsf{T} \ . \tag{8}$$

All 84 subvectors are concatenated into a joint false-color feature vector $\mathbf{F} \in \mathbb{R}^{252}$:

$$\mathbf{F} = [\mathbf{f}_1^\mathsf{T}, \mathbf{f}_2^\mathsf{T}, ..., \mathbf{f}_3^\mathsf{T}] \ . \tag{9}$$

Finally, the complete 1D feature vector $\mathbf{x}$ utilized as the input for the boosting classifiers is formed by a flat concatenation of all subvectors:

$$\mathbf{x} = [\mathbf{R}^\mathsf{T}, \Delta^\mathsf{T}, \mathbf{Q}^\mathsf{T}, \mathbf{F}^\mathsf{T}] \, . \tag{10}$$

This results in a total of $9 + 8 + 8 + 252 = 277$ dimensions per pixel capturing absolute spectral information, local gradients, nonlinear multi-band relationships, and false-color combinations.

### 4.2 Optimal filter selection

The introduced multispectral MHP database provides spectral data of 24 wavelengths ranging from 800 nm to 1600 nm. In practice, capturing all 24 bands simultaneously is rarely feasible due to hardware and economic limitations. The employed camera array only supports a limited number of sensors, and adjacent spectral bands frequently contain highly correlated information. To reduce redundancy and minimize hardware and processing requirements, the Conditional Filter Band Selection (CFBS) algorithm [23] was applied. This method identifies a minimal subset of wavelengths that retain the most relevant spectral information for the classification task. The resulting filter set is the optimal trade-off between system simplicity and classification capability. Although the precise subset is tailored to the given material types and measurement setup, the underlying approach is general and can be transferred to other multispectral systems and application domains. The specific configuration used in this work, as well as its influence on classification accuracy, will be described in detail in the subsequent results Sect. 5.1.

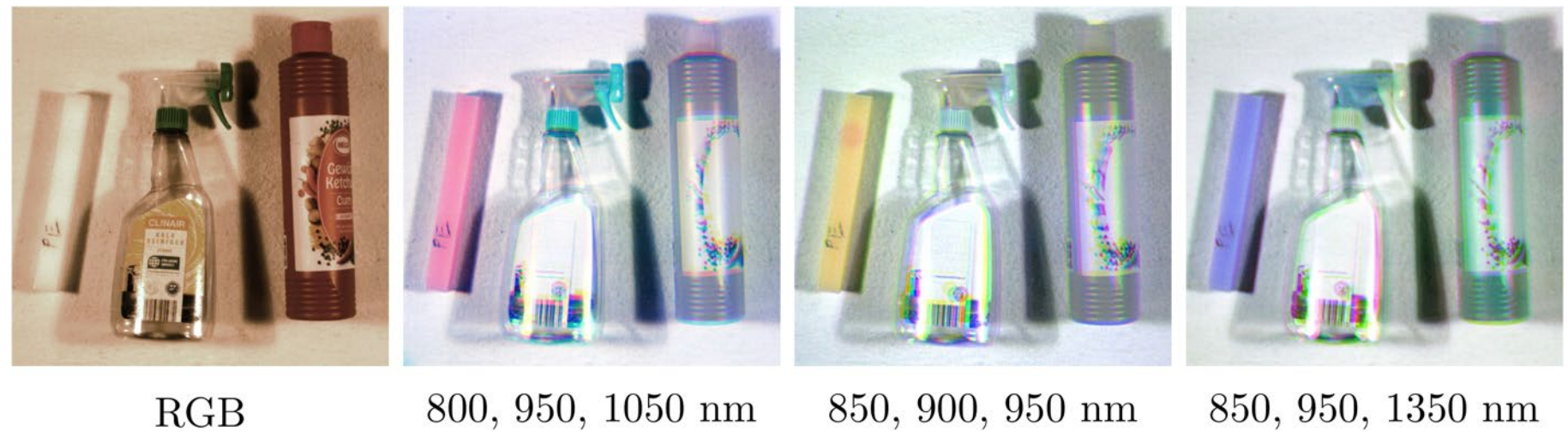

**Fig. 6** Original RGB image and false-color images after superimposition of three different infrared channels as Red, Green, and Blue. Different sorts of plastic material appear in different colors in the IR and, hence, can be distinguished

### 4.3 Segmentation and classification models

To enable spatially resolved material identification, the classification task is formulated as a pixel-wise classification, where each pixel in the multispectral image is assigned a label corresponding to a specific plastic type (HDPE, LDPE, PET, PP, PS, PTFE, and PVC) or the auxiliary class “unknown" to handle unidentifiable materials.

The classification is applied exclusively within object regions. These regions are extracted using a YOLOv11 [32] object detection model, which was retrained on the RGB images of the HPM dataset to generate precise binary segmentation masks. These masks are subsequently projected onto the corresponding multispectral data, ensuring that the background is excluded, and each object is processed independently across all spectral bands. Because of the parallax shifts generated by the physical baselines of the camera array, projecting the RGB-derived binary mask onto the peripheral IR channels inevitably introduces slight spatial misalignments at the object boundaries. However, the subsequent object-level majority voting mechanism, which is explained in detail in Sect. 4.4, successfully acts as a robust filter against these uncertain boundary pixels, ensuring that minor edge artifacts generated by the mask shift do not compromise the final material classification.

Within these segmented regions, a pixel-wise classification is performed. To determine the most effective algorithmic approach, a diverse set of machine learning models was experimentally validated. Random forest [33] and k-nearest neighbor [34] algorithms demonstrated substantially lower classification performance in preliminary tests and were, therefore, excluded from further investigation. Partial Least Squares Discriminant Analysis [35] (PLS-DA) and 1D Convolutional Neural Networks (1D-CNNs) [36] were implemented to serve as established baselines for spectral signature analysis. However, following our feature extraction and optimal filter selection, the classification problem is reduced to a low-dimensional format. For this data format, the boosting-based classifiers consistently outperformed the linear PLS-DA method, while requiring minimal architectural tuning and computational resources compared to 1D-CNNs [37]. Consequently, to meet the strict constraints of real-time industrial sorting applications, the boosting-based approach was selected for the final classification pipeline.

Among boosting methods, several popular implementations were explored. Gradient Boosting (GB) [38] constructs additive models by sequentially fitting shallow decision trees to the residual of previous learners, with performance strongly influenced by hyperparameters such as learning rate, depth, and subsampling. Extreme Gradient Boosting (XGBoost) [39] extends GB with regularization terms that control model complexity, improving scalability, and reducing overfitting. LightGBM [40] optimizes training speed and memory usage through histogram-based split finding, gradient-based sampling, and feature bundling, making it particularly efficient for large datasets.

CatBoost [41], in turn, avoids prediction shift through ordered boosting and uses symmetric tree structures, offering robust generalization and efficient inference.

### 4.4 Model training and uncertainty quantification

To maximize the performance and robustness of the selected boosting algorithms, an exhaustive grid search [42] was employed for hyperparameter optimization on the training and validation splits of the MHP database. The search space encompassed learning rates between 0.05 and 0.2, maximum tree depths between 7 and 12, and the number of estimators (iterations) ranging from 300 to 1100.

To further enhance generalization and mitigate overfitting, a bagging-based ensemble strategy [43] was implemented for each classifier type. Instead of relying on a single predictor, an ensemble of ten independent models was trained in parallel. Each member of the ensemble was trained on a randomized subset of the training data, generated via bootstrap sampling.

During inference, this ensemble approach allows us to estimate prediction confidence, as uncertainty arises from both inherent sensor noise and model limitations [44]. For every pixel, the ensemble provides a distribution of independent predictions. We determine the final class probability by computing the mean across all ten models, alongside the standard deviation. A pixel's classification is deemed uncertain if its standard deviation exceeds 20% or if its mean predicted probability falls below 95%. Pixels that fail to meet these conservative thresholds are assigned the auxiliary label “unknown." While this 95% threshold was deliberately chosen to prioritize high classification certainty and strictly minimize false positives, it represents an adjustable parameter that can be flexibly adapted to the specific operational trade-off between recovery rate and required material purity.

Finally, to transition from pixel-level predictions to object-level sorting decisions, a spatial majority voting scheme is applied within the boundaries defined by the YOLOv11 segmentation masks. All valid, confident pixels within a given mask vote for a material class. By completely excluding uncertain pixels from this voting process, the most frequently predicted label determines the global material classification of the entire object with high reliability, and unprecise decisions at object borders become negligible.

### 4.5 Experimental setup and evaluation strategy

The complete Multispectral Household Plastic Classification (MHPC) workflow is summarized in Fig. 7. During the model development phase, the MHP database was partitioned into a dedicated training and validation split (85% and 15%, respectively) to select the best-performing configuration. The resulting model was evaluated on an independent test set.

First, the raw multispectral images undergo preprocessing to correct geometric misalignments, apply color correction, and normalize intensity values. Second, we extract discriminative features including characteristic spectral fingerprints, spectral slopes, and false-color composites from the preprocessed images. These features serve as input to the classification algorithm. The models are trained on the 85% training partition and validated on the held-out 15% validation split to find the best hyperparameter combination and the best-performing model.

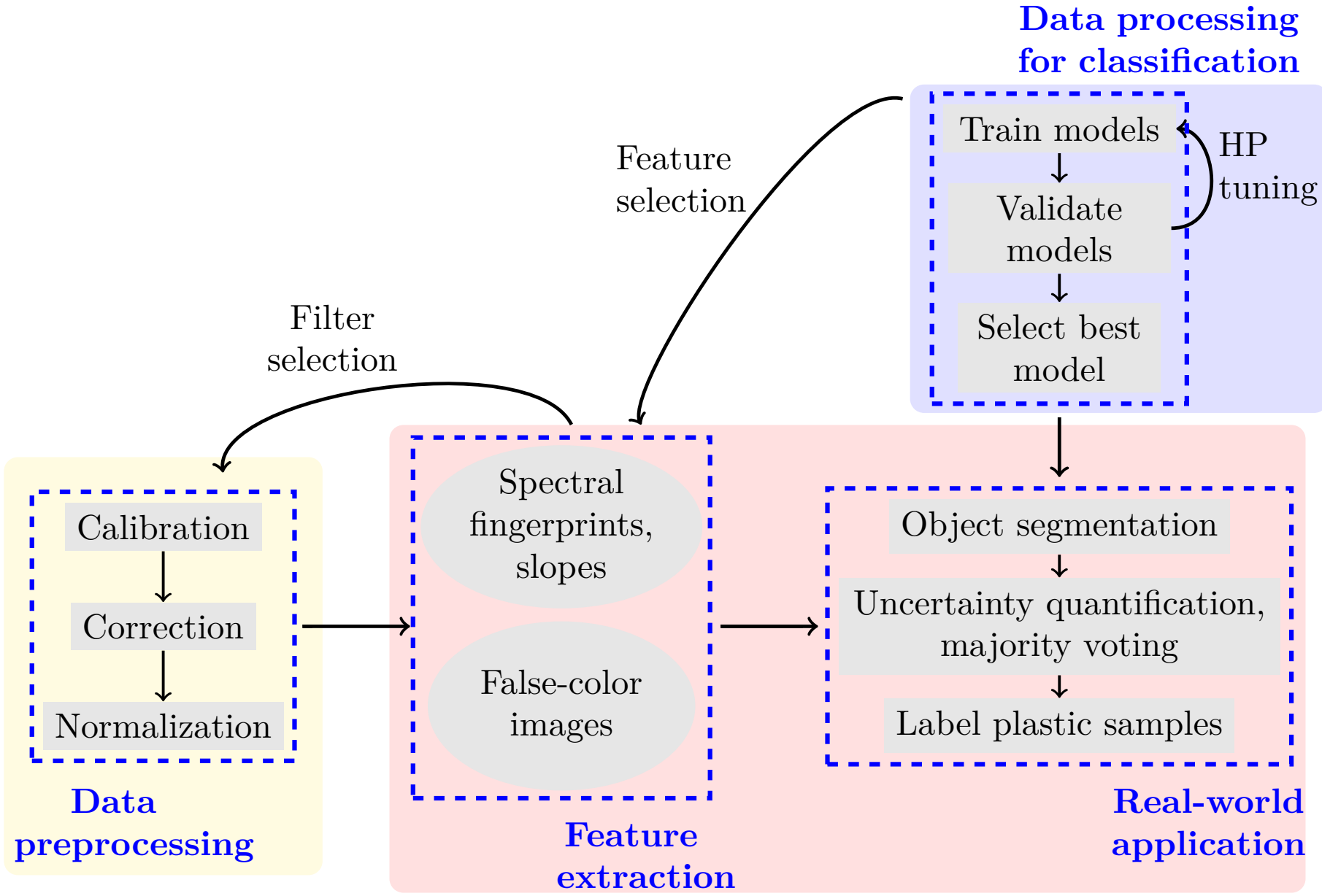

**Fig. 7** Classification pipeline. The model training (blue) only has to be performed once. The preprocessing steps (yellow) only have to be performed once after starting the application, while the rest of the processing (red) has to be performed for each recording

To ensure realistic performance estimation and true generalization, the final evaluation was conducted on a completely independent, external dataset. This external test set consists of physical plastic objects representing an operational industrial setting with controlled IR illumination. Crucially, these samples were not part of the MHP database and were strictly excluded from all prior training and tuning phases.

Finally, the quantitative performance of the classification algorithms on this independent test set is evaluated primarily using normalized confusion matrices. These matrices illustrate the proportion of predictions for each class relative to the total number of samples, allowing for a detailed analysis of per-class accuracies, off-diagonal misclassification trends, and overall algorithmic robustness.

## 5 Results

### 5.1 Filter selection evaluation

Before comparing classifier performances, CFBS was applied to determine the optimal filter subset. Given the MHP dataset comprises 24 wavelengths ranging from 800 nm to 1600 nm, and the system configuration includes eight grayscale cameras covering the 800–1150 nm range and one SWIR camera extending to 1600 nm, a comprehensive evaluation of all feasible filter combinations was conducted. At low filter counts, classification accuracies varied strongly depending on the selected bands, with minima around 1–2% and maxima up to 87%. As the number of filters increased, the variability decreased and accuracies improved, peaking around eight filters, where the maximal accuracy approached almost 87%. Figure 8 visualizes this empirical dependence between filter count and classification accuracy, indicating an optimal point beyond which additional filters yield diminishing or even degrading performance due to noise accumulation, spectral redundancy, and overfitting.

Based on this ablation analysis, we selected nine filters 800 nm, 850 nm, 900 nm, 950 nm, 975 nm, 1000 nm, 1050 nm, 1100 nm, and 1350 nm. While employing multiple SWIR cameras covering the 1200–1600 nm region increased classification accuracy by 3.5–7%, this configuration was not pursued further due to its higher cost. The current setup, therefore, uses a single SWIR camera. Nevertheless, once SWIR technology becomes more affordable, the system can be easily extended to integrate multiple SWIR channels without significant architectural modifications.

### 5.2 Feature importance analysis

The final feature vector comprises 277 dimensions, 252 of which are derived from three-channel false-color combinations. To evaluate the impact of potential collinearity introduced by this high-dimensional space and to provide a physical and algorithmic justification for the model's decisions, a feature importance analysis was conducted for the best-performing LightGBM model.

Figure 9 on the left illustrates the top 20 features based on the native LightGBM information gain metric [38], which measures the total reduction of impurity across all trees. The analysis reveals that the model primarily relies on raw spectral intensities (e.g., WL_1125 nm and WL_900 nm) and mathematically derived slopes (e.g., Slope_1050–1125 nm). Notably, not a single false-color triplet combination appears among the top features. This quantitatively demonstrates that the tree-based LightGBM algorithm successfully handles the introduced collinearity by bypassing redundant derived features.

To further investigate the decision-making process at the prediction level, a Shapley Additive exPlanations (SHAP) [45] interaction analysis was performed, where the results

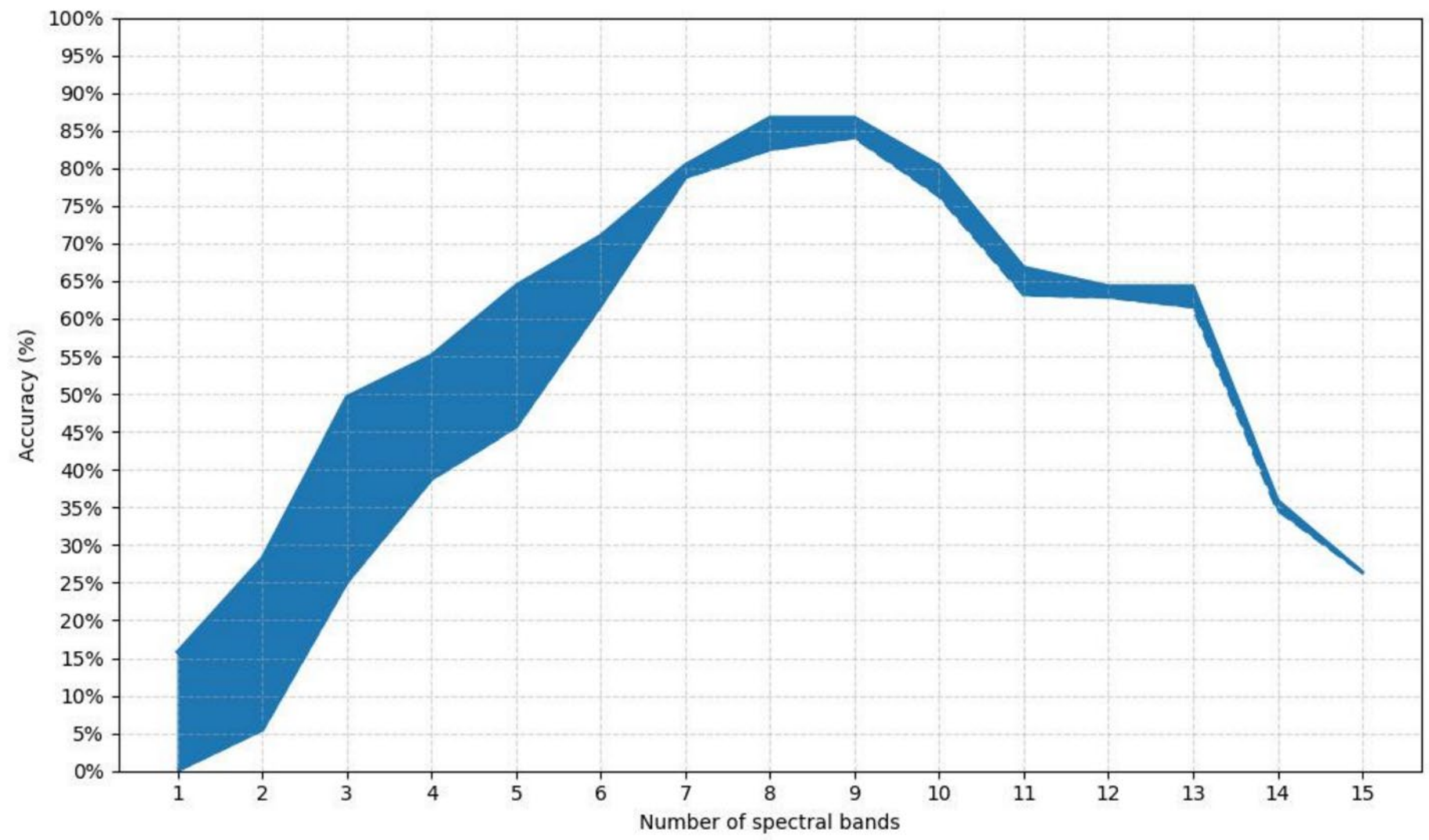

**Fig. 8** Impact of spectral filter count on classification accuracy. Initially, with fewer filters, multiple combinations exist, leading to a broader range of accuracy outcomes. As the number of filters increases, the variability decreases, and the accuracy stabilizes. Upon selecting all available filters, only one combination remains, resulting in a fixed accuracy value. Beyond an optimal point, adding more filters may degrade performance due to increased noise, spectral redundancy, and overfitting

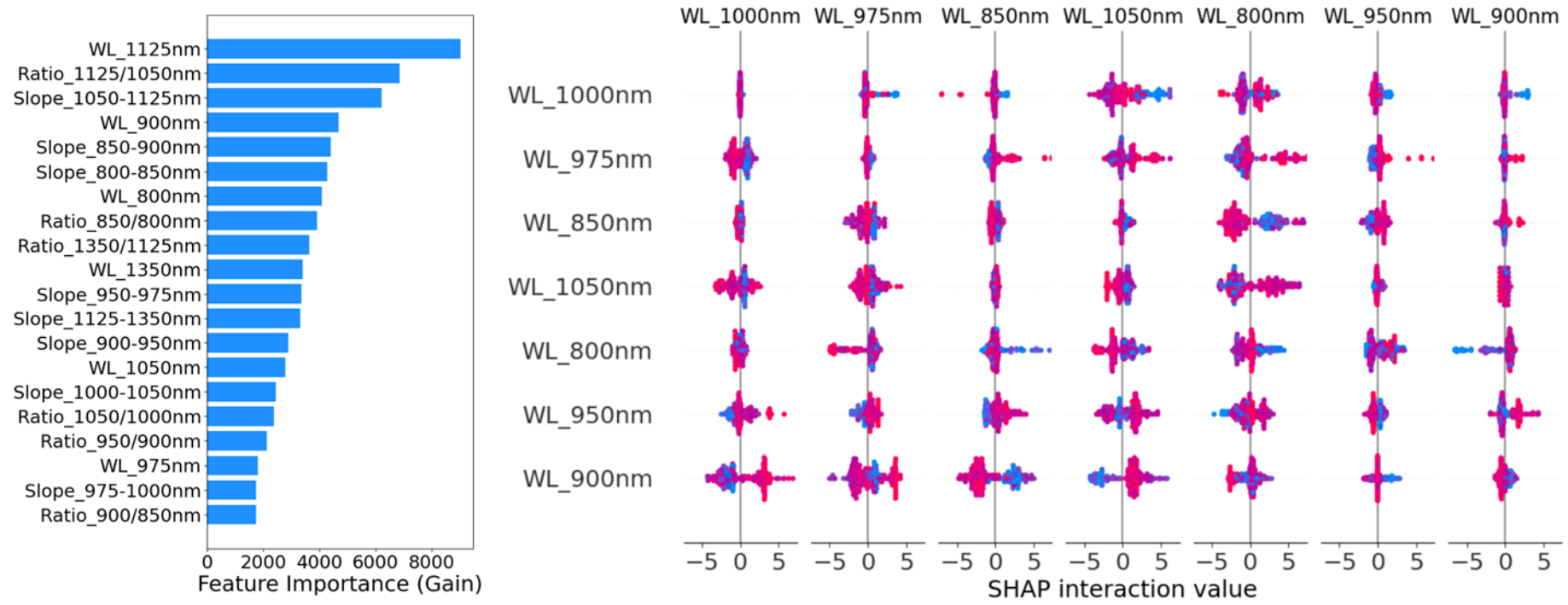

**Fig. 9** Feature Importance Analysis for the LightGBM model. Left: Top 20 features based on the native information gain metric, demonstrating the algorithmic efficiency of slopes. Right: SHAP interaction value matrix illustrating the prediction-level reliance on the interplay of raw spectral intensity bands

are shown in Fig. 9 on the right. The SHAP matrix confirms that the final predictions are fundamentally driven by the interplay of the raw intensity bands rather than the derived combinations. The strong dominance of features around the 1050 nm and 1125 nm wavelengths provides a physical justification for the classification, as this specific near-infrared region captures characteristic absorption variations that are highly discriminative for common household polymers such as PET, PS, and PP, and further validates and confirms our filter selection.

Overall, the raw spectral bands contain the fundamental physical information needed for the classification. The calculated spectral slopes merely make it easier for the model to find the best decision boundaries during training, improving algorithmic efficiency without losing the physical context.

### 5.3 Classification performance

Using the optimized nine wavelengths and the features, the pixel-wise classification performance of the four boosting classifiers was evaluated on the independent test set consisting of 28 post-consumer items (4 per class). The resulting normalized confusion matrices are shown in Fig. 10.

Each classifier is able to recognize HDPE, PET, PP, PTFE, and PVC with an accuracy of at least 80%. However, depending on the model, even higher accuracies can be reached for several materials. Notably, PTFE is classified with an accuracy close to 100% for most of the classifiers. This is attributed to its very constant spectral response at higher wavelengths, especially beyond 1200 nm, which is unique among the tested plastics. In contrast, PS consistently yields the lowest classification accuracy. This may be due to its high variance at longer wavelengths and almost no variance at the 1150 nm drop. This is confirmed by the models' application outputs in Fig. 11, where PS is more frequently misclassified than other materials across the models. The labels were verified by referencing the RICs on the plastic items.

Among the evaluated classifiers, LightGBM achieved the highest overall accuracy of 86.70% and the fastest per-pixel runtime of 2.603 $\mu$s, as summarized in Table 4. In the setup shown in Fig. 11, all plastics were classified correctly except for a single PS sample that was misclassified. XGBoost reached 85.57% accuracy with particularly strong results for PP, PS, and HDPE, where the precision exceeded 85%.

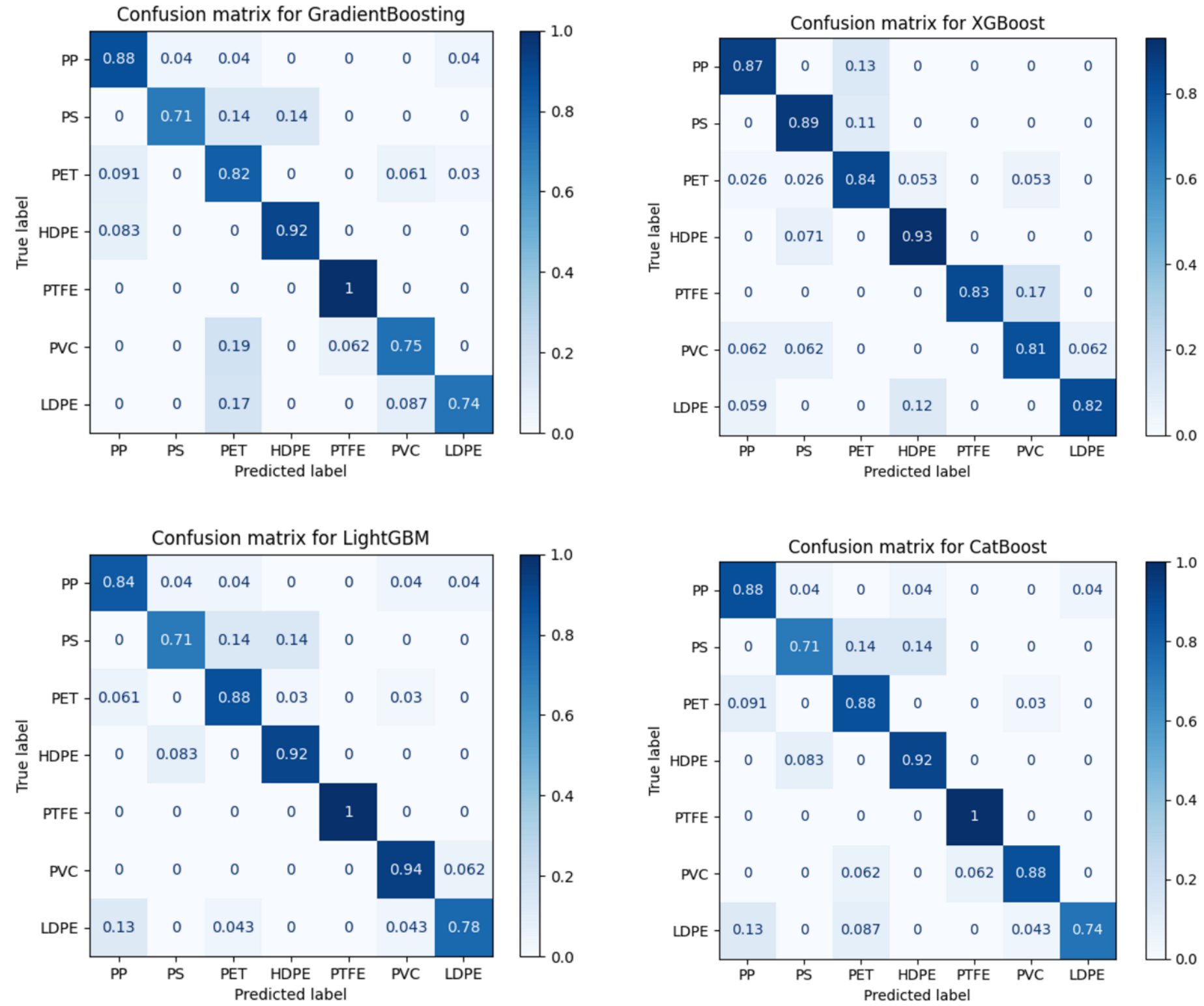

**Fig. 10** Normalized confusion matrices for GB (top left), XGBoost (top right), LightGBM (bottom left), and CatBoost (bottom right). Results are shown rounded to two decimal points

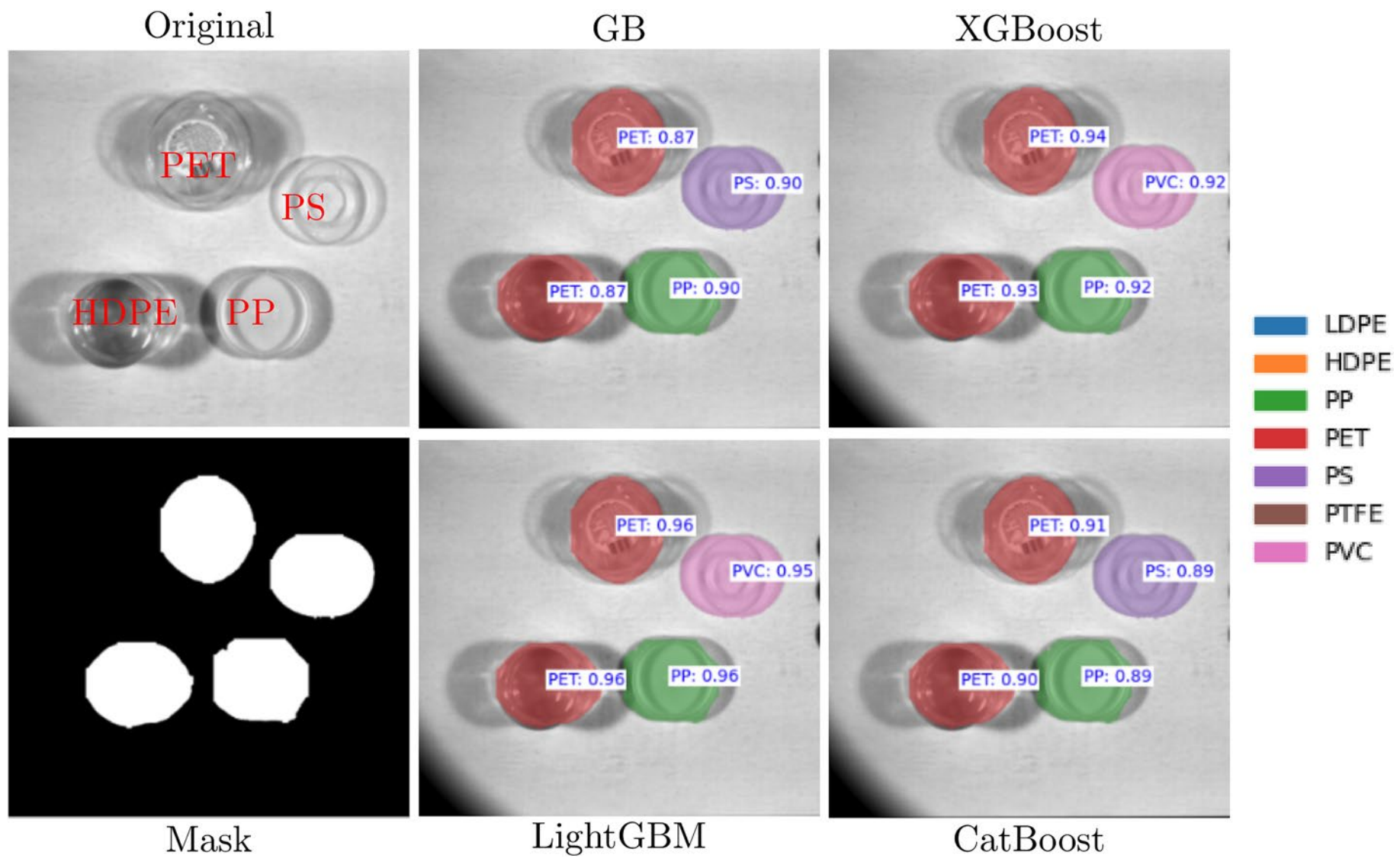

**Fig. 11** Results and classification accuracy of the four different classifiers, the ground truth, and the mask used for classification. The ground truth labels were manually obtained based on the RICs marked on each plastic sample. On the top left, one grayscale channel is shown with the correct labels of the plastic materials. For GB and CatBoost, all plastics are correctly labeled

**Table 4** Per-pixel runtime, classification throughput (decisions per second), and overall performance metrics (accuracy, weighted precision, weighted recall, and weighted F1) of GB, XGBoost, LightGBM, and CatBoost

| | GB | XGBoost | LightGBM | CatBoost |
|---|---|---|---|---|
| Runtime ($\mu$s) | 8.658 | 27.533 | 2.603 | 12.586 |
| Decisions per second | 115 500 | 36 300 | 384 300 | 79 500 |
| Accuracy | 83.14% | 85.57% | 86.70% | 85.85% |
| Weighted precision | 86.54% | 86.22% | 86.59% | 83.91% |
| Weighted recall | 83.14% | 85.86% | 86.71% | 86.79% |
| Weighted F1 | 86.03% | 85.73% | 85.97% | 84.94% |

**Table 5** Comparison of our best performing LightGBM model with PLS-DA [35] and 1D-CNN [36]

| | PLS-DA [35] | 1D-CNN [36] | LightGBM [40] |
|---|---|---|---|
| Accuracy | 53.86% | 71.06% | **86.70%** |

Best results is shown in bold

The confusion matrix reveals that most misclassifications occurred for PVC, which was occasionally predicted as PP or PS. In the application example of Fig. 11, the PS sample was also incorrectly labeled as PVC. With a per-pixel runtime of 27.533 $\mu$s, XGBoost is considerably slower than LightGBM. Gradient Boosting achieved 83.14% overall accuracy and maintained at least 80% for PP, PET, HDPE, and PTFE, but showed weaker performance for PS, PVC, and LDPE. Despite its relatively fast runtime of 8.658 $\mu$s per pixel, its robustness across classes is limited. CatBoost reached 85.85% accuracy and outperformed GB and XGBoost for PP and PVC, though it required 12.586 $\mu$s per pixel. In the application example, all plastics were correctly labeled.

Given the class imbalance in the HPM dataset, relying solely on overall accuracy could mask underperformance on minority classes such as PS or PVC, which comprise six samples. Therefore, Table 4 also includes weighted precision, weighted recall, and weighted F1-scores for all classifiers. By weighting these metrics based on the number of physical samples per material class, the evaluation explicitly accounts for imbalances. The high weighted F1-score of 85.97% for LightGBM confirms the model's robustness and ensures that minority classes are identified with comparable reliability to majority classes, as the score aligns closely with the overall accuracy of 86.70%. The other models demonstrate similarly robust behavior, with their weighted F1-scores closely tracking or even exceeding their overall accuracies.

To summarize, the results presented in Table 4 and Fig. 11 demonstrate that LightGBM provides the most favorable trade-off between accuracy and runtime efficiency, while the other classifiers offer competitive performance with slightly longer processing times or lower robustness for certain plastics. The example in Fig. 11 further confirms the transferability of the models from the laboratory setup to real application scenarios.

### 5.4 Comparison with the literature

To contextualize the performance of the boosting classifiers, they were also evaluated on the independent test set against PLS-DA [35] and a 1D-CNN [36] trained on the exact same feature set. The results are given in Table 5. PLS-DA achieved an overall classification accuracy of only 53.86%. The 1D-CNN achieved an accuracy of 71.06%. The substantial performance gap between these baselines and the boosting classifiers empirically validates our choice of tree-based boosting methods for low-dimensional spectral data.

Furthermore, Table 6 compares our proposed LightGBM approach against existing standard and low-cost spectral methods. Standard point-measurement NIR systems [13, 14] typically achieve very high classification accuracies exceeding 95%, but rely on expensive, specialized hardware and can usually only process one plastic object at a time. While previous low-cost sensor systems [20, 21] are restricted to classifying up to six plastic types with average accuracies ranging from 62.08% to 90.0%, our multispectral camera array expands the classification scope to the seven most common household plastics while maintaining a highly competitive overall accuracy of 86.70% using off-the-shelf components.

### 5.5 Limitations

Despite demonstrating robust performance in an uncalibrated industrial test setting, the proposed system exhibits several operational limitations.

First, the current model cannot reliably identify out-of-distribution materials. Plastic types and substances that were not part of the training data, such as engineering plastics like polyoxymethylene (POM) or acrylonitrile butadiene styrene (ABS), may share localized spectral characteristics with known plastic and cause false-positive classifications. While such materials are statistically rare in standard municipal waste streams, adapting the system to specialized recycling sectors requires retraining of the classifiers on an extended training dataset.

Second, even though the model performs well, certain plastics remain difficult to distinguish. Black plastics, for instance, absorb most of the infrared light and thus exhibit low contrast in the images, making classification less reliable. Similarly, extremely thin, transparent plastic films are hard to classify. While thicker transparent components can still be identified reliably, thin films pose a challenge due to their low signal intensity and partial visibility of the background. In contrast, printed or colored plastics are handled well by the system, as dyes and surface features do not substantially alter the material's underlying chemical composition.

**Table 6** Comparison of the proposed method with existing spectral and optical sorting technologies

| Reference | Method/Technology | Throughput | #Plastic types | Accuracy |
|---|---|---|---|---|
| Chen et al. [13], Maier et al. [14] | NIR Spectrometer + Chemometrics | Low | ~10 | >95% |
| West et al. [20] | Multispectral NIR sensor + DL | Moderate | 6 | 62.08% |
| Werner et al. [21] | Multispectral NIR sensor + KNN | Moderate | 5 | ~90.0% |
| **Proposed** | **NIR camera array + LightGBM** | **High** | **7** | **86.70%** |

Best results is shown in bold

Furthermore, the hyperparameter tuning in this work utilized a random 85/15 split of the training data. Given the limited number of physical samples per class, this random split during the Grid Search phase poses a methodological risk of overfitting to specific physical samples. However, it must be emphasized that the final reported classification accuracy was evaluated on completely distinct, unseen physical objects using additional images. The consistently high performance on this independent test set demonstrates that the models successfully learned generalizable material features.

## 6 Conclusion

In this work, a classification model for identifying different household plastics based on multispectral imaging data was developed and evaluated. To this end, a multispectral household plastic dataset was introduced. Four different machine learning classifiers were trained on the recorded dataset and evaluated both on the dataset and on new unseen plastic objects. Their individual predictions were improved using an ensemble strategy, which enhanced the robustness of the system and led to an overall improvement in classification performance compared to the standalone models. The classifiers achieved accuracy levels between 84% and 87% on the unseen objects that were not part of the training data, outperforming existing methods that report between 62% and 90% accuracy for only five to six plastic types. Our approach attains 87% across seven types, demonstrating improved robustness and generalization. The fast processing time enables frame-wise classification at the scale required for conveyor-belt speeds in automated sorting facilities. The hardware is entirely built from off-the-shelf components and its modular design permits the replacement of individual cameras or filters in case of malfunction and allows extensions, for example, by adding further wavelength channels. This makes the system technically adaptable to different plastic classes or changing operational conditions without requiring a complete redesign.

Future developments will aim to expand the classification scope beyond the seven most common household plastics. In particular, we intend to incorporate additional plastic types and introduce the ability to detect and differentiate non-plastic materials such as paper, textiles, wood, and metals. Additionally, to improve the statistical validity of the hyperparameter optimization process when working with limited physical samples, investigating object-wise split strategies, such as k-fold cross-validation or Leave-One-Object-Out approach, will be part of future research. These advancements will significantly broaden the applicability and robustness of the system in complex material streams and further enhance its utility in waste management scenarios.

**Acknowledgements**

The authors thank the Sielaff GmbH & Co. KG for providing plastic samples.

**Author contributions**

KK is a graduate student and received the master's degree in medical engineering with a focus on medical electronics from Friedrich-Alexander Universität Erlangen-Nürnberg (FAU), Germany, in 2022. Since 2023 KK has been pursuing the Ph.D. degree with the Chair of Multimedia Communications and Signal Processing. KK's research includes image and video signal processing, and multispectral imaging. In 2024, KK won the FAU sustainability award with the work toward multispectral plastic classification. JS received the diploma degree in electrical engineering, electronics, and information technology and the Ph.D. and Habilitation degrees from the Chair of Multimedia Communications and Signal Processing, Friedrich-Alexander Universität Erlangen-Nürnberg, Germany, in 2006, 2011, and 2018, respectively. JS is currently a senior scientist and a lecturer with the Chair of Multimedia Communications and Signal Processing, Friedrich-Alexander Universität Erlangen-Nürnberg. JS has authored or coauthored more than 100 technical publications. His research interests include image and video signal processing, signal reconstruction and coding, signal transforms, and linear systems theory. JS received the Dissertation Award of the Information Technology Society of the German Electrical Engineering Association as well as the Dissertation Award of the Staedtler-Foundation, both in 2012. In 2007, JS received diploma awards from the Institute of Electrical Engineering, Electronics, and Information Technology, Erlangen, as well as from the German Electrical Engineering Association. JS also received scholarships from the German National Academic Foundation and the Lucent Technologies Foundation. JS was a co-recipient of four best paper awards. AK received the Dipl.-Ing. and Dr.-Ing. degrees in electrical engineering from RWTH Aachen University, Aachen, Germany, in 1989 and 1995, respectively. AK joined Siemens Corporate Technology, Munich, Germany, in 1995, and became the Head of the Mobile Applications and Services Group in 1999. Since 2001, AK has been a Full Professor and the Head of the Chair of Multimedia Communications and Signal Processing, Friedrich-Alexander University Erlangen-Nürnberg (FAU), Germany. From 1997 to 2001, AK was the Head of the German MPEG delegation. From 2005 to 2007, AK was a Vice Speaker of the DFG Collaborative Research Center 603. From 2015 to 2017, AK has served as the Head of the Department of Electrical Engineering and the Vice Dean of the Faculty of Engineering, FAU. AK has authored around 500 journal and conference papers and has over 120 patents granted or pending. AK's research interests include image and video signal processing and coding and multimedia communication. AK is a member of the IEEE Image, Video, and Multidimensional Signal Processing Technical Committee, the Scientific Advisory Board of the German VDE/ITG, the Bavarian Academy of Sciences and Humanities, and the European Academy of Sciences and Arts. AK is a member of the Editorial Board of the *IEEE Circuits and Systems Magazine*. He was the Siemens Inventor of the Year 1998 and received the 1999 ITG Award. AK received several IEEE best paper awards, including the Paul Dan Cristea Special Award in 2013 and AK's group won the Grand Video Compression Challenge from the Picture Coding Symposium 2013. The Faculty of Engineering, FAU, and the State of Bavaria honored AK with teaching awards, in 2015 and 2020, respectively. He served as an Associate Editor for IEEE Transactions on Circuits and Systems for Video Technology. AK was a Guest Editor for the IEEE Journal of Selected Topics in Signal Processing.

**Funding**
Open Access funding enabled and organized by Projekt DEAL. The authors gratefully acknowledge that this work has been supported by the Bayerische Transformations- und Forschungsstiftung (Bavarian Transformation and Research Foundation) under project number AZ-1547-22.

**Data availability**
To access the dataset please use the following link: https://github.com/FAU-LMS/MHPM.

## Declaration

**Conflict of interest**
The authors declare no conflict of interest.

**References**

1. Plastics - The Facts. https://plasticseurope.org/knowledge-hub/plastics-the-facts-2022-2/
2. J.R. Jambeck, R. Geyer, C. Wilcox, T.R. Siegler, M. Perryman, A. Andrady, R. Narayan, K.L. Law, Plastic waste inputs from land into the ocean. Science **347**(6223), 768–771 (2015)
3. A.S. Pottinger, R. Geyer, N. Biyani, C.C. Martinez, N. Nathan, M.R. Morse, C. Liu, S. Hu, M. Bruyn, C. Boettiger et al., Pathways to reduce global plastic waste mismanagement and greenhouse gas emissions by 2050. Science **386**(6726), 1168–1173 (2024)
4. Bashmakov, I.A., Nilsson, L.J., Acquaye, A., Bataille, C., Cullen, J.M., Fischedick, M., Geng, Y., Tanaka, K., et al.: Climate change 2022: Mitigation of climate change. Contribution of Working Group III to the Sixth Assessment Report of the Intergovernmental Panel on Climate Change, Chapter 11 (2022)
5. A.D. Macheca, B. Mutuma, J.L. Adalima, E. Midheme, L.H. Lúcas, V.K. Ochanda, S.D. Mhlanga, Perspectives on plastic waste management: challenges and possible solutions to ensure its sustainable use. Recycling **9**(5), 1–66 (2024)
6. Kumar, R., Verma, A., Shome, A., Sinha, R., Sinha, S., Jha, P.K., Kumar, R., Kumar, P., Shubham, Das, S., et al.: Impacts of plastic pollution on ecosystem services, sustainable development goals, and need to focus on circular economy and policy interventions. Sustainability 13(17), 9963 (2021)
7. Wolberg, C.: The Performance of Recycled Plastics Vs Virgin Plastics. https://oceanworks.co/blogs/ocean-plastic-news/the-performance-of-recycled-vs-virgin-plastics
8. ASTM International, *Standard Practice for Coding Plastic Manufactured Articles for Resin Identification* (ASTM International, West Conshohocken, PA, 2013)
9. Cantner, J., Gerstmayr, B., Pitschke, T., Tronecker, D., Hartleitner, B., Kreibe, S.: Bewertung der Verpackungsverordnung – Evaluierung der Pfandpflicht. Technical Report UBA-FB 001363/1, Umweltbundesamt (2010). last accessed: 21-05-2025. https://www.umweltbundesamt.de/sites/default/files/medien/461/publikationen/3932.pdf
10. S. Serranti, G. Bonifazi, Techniques for separation of plastic wastes, in *Use of Recycled Plastics in Eco-efficient Concrete*. (Elsevier, Amsterdam, 2019), pp.9–37
11. O. Kökkılıç, S. Mohammadi-Jam, P. Chu, C. Marion, Y. Yang, K.E. Waters, Separation of plastic wastes using froth flotation-an overview. Adv. Coll. Interface. Sci. **308**, 102769 (2022)
12. A. Dolet, F. Varray, S. Mure, T. Grenier, Y. Liu, Z. Yuan, P. Tortoli, D. Vray, Spatial and spectral regularization to discriminate tissues using multispectral photoacoustic imaging. EURASIP J. Adv Signal Process. **2018**(1), 39 (2018)
13. X. Chen, J. Zhou, L.-M. Yuan, G. Huang, X. Chen, W. Shi, Spectroscopic identification of environmental microplastics. IEEE Access **9**, 47615–47620 (2021)

14. G. Maier, R. Gruna, T. Längle, J. Beyerer, A survey of the state of the art in sensor-based sorting technology and research. IEEE Access **12**, 6473–6493 (2024)
15. I. Cortesi, A. Masiero, G. Tucci, K. Topouzelis, UAV-based river plastic detection with a multispectral camera. Int. Arch. Photogramm. Remote. Sens. Spat. Inf. Sci. **43**, 855–861 (2022)
16. M.M. Duarte, L. Azevedo, Automatic detection and identification of floating marine debris using multispectral satellite imagery. IEEE Trans. Geosci. Remote Sens. **61**, 1–15 (2023)
17. Y. Jiang, C. Li, Detection and discrimination of cotton foreign matter using push-broom based hyperspectral imaging: System design and capability. PLoS ONE **10**(3), 1–18 (2015)
18. G. Bonifazi, G. Capobianco, S. Serranti, Fast and effective classification of plastic waste by pushbroom hyperspectral sensor coupled with hierarchical modelling and variable selection. Resour. Conserv. Recycl. **197**, 107068 (2023)
19. A. Picon, O. Ghita, S. Rodriguez-Vaamonde, P.M. Iriondo, P.F. Whelan, Biologically-inspired data decorrelation for hyper-spectral imaging. EURASIP J. Adv. Signal Process. **2011**(1), 66 (2011)
20. G. West, T. Assaf, U. Martinez-Hernandez, Towards low-cost plastic recognition using machine learning and multi-spectra near-infrared sensor. IEEE SENSORS, 1–4 (2023)
21. T. Werner, M. Dawoud, D. Aschenbrenner, I. Taha, Cost-efficient detection of plastics from post-consumer packaging waste using selected bands in the near-infrared spectrum. Macromol. Mater. Eng. **310**(10), e00143 (2025)
22. N. Genser, J. Seiler, A. Kaup, Camera array for multi-spectral imaging. IEEE Trans. Image Process. **29**, 9234–9249 (2020)
23. K. Kossira, D. Schön, J. Seiler, A. Kaup, Conditional optimal filter selection for multispectral object classification, in *2024 IEEE International Conference on Image Processing (ICIP)*. (IEEE, 2024), pp.2121–2127
24. T. Dang, C. Hoffmann, C. Stiller, Continuous stereo self-calibration by camera parameter tracking. IEEE Trans. Image Process. **18**(7), 1536–1550 (2009)
25. T. Rahman, N. Krouglicof, An efficient camera calibration technique offering robustness and accuracy over a wide range of lens distortion. IEEE Trans. Image Process. **21**(2), 626–637 (2011)
26. S. Liu, M. Liu, Z. Yang, An image auto-focusing algorithm for industrial image measurement. EURASIP J. Adv. Signal Process. **2016**(1), 70 (2016)
27. F. Sippel, J. Seiler, A. Kaup, High-resolution hyperspectral video imaging using a hexagonal camera array. J. Opt. Soc. Am. A **41**(12), 2303–2315 (2024)
28. F. Sippel, J. Seiler, A. Kaup, Cross spectral image reconstruction using a deep guided neural network, in *2023 IEEE International Conference on Image Processing (ICIP)*. (2023), pp.226–230
29. X. Ding, L. Hu, S. Zhou, X. Wang, Y. Li, T. Han, D. Lu, G. Che, Snapshot depth-spectral imaging based on image mapping and light field. EURASIP J. Adv. Signal Process. **2023**(1), 24 (2023)
30. K. Kossira, J. Seiler, A. Kaup, Inter-camera color correction for multispectral imaging with camera arrays using a consensus image, in *2024 IEEE 26th International Workshop on Multimedia Signal Processing (MMSP)*. (IEEE, 2024), pp.1–6
31. C. Saranya, G. Manikandan, A study on normalization techniques for privacy preserving data mining. Int. J. Eng. Technol.(IJET) **5**(3), 2701–2704 (2013)
32. G. Jocher, J. Qiu, A. Chaurasia, Ultralytics YOLO11. GitHub repository (2024).
33. L. Breiman, Random forests. Mach. Learn. **45**, 5–32 (2001)
34. J. Laaksonen, E. Oja, Classification with learning k-nearest neighbors, in *Proceedings of International Conference on Neural Networks (ICNN'96)*, vol. 3, (IEEE, 1996), pp. 1480–1483
35. Sjöström, M., Wold, S., Söderström, B.: PLS discriminant plots. In: Pattern Recognition in Practice, pp. 461–470. Elsevier, (1986)
36. P. Bian, W. Li, Y. Jin, R. Zhi, Ensemble feature learning for material recognition with convolutional neural networks. EURASIP J. Image and Video Process. **2018**(1), 64 (2018)
37. H. Xu, Z. Han, S. Feng, H. Zhou, Y. Fang, Foreign object debris material recognition based on convolutional neural networks. EURASIP J. Image and Video Process. **2018**(1), 21 (2018)
38. J.H. Friedman, Greedy function approximation: a gradient boosting machine. Ann. Stat., 1189–1232 (2001)
39. T. Chen, C. Guestrin, XGBoost: A scalable tree boosting system, in *Proceedings of the 22nd ACM SIGKDD International Conference on Knowledge Discovery and Data Mining*. (2016), pp. 785–794
40. G. Ke, Q. Meng, T. Finley, T. Wang, W. Chen, W. Ma, Q. Ye, T.-Y. Liu, LightGBM: a highly efficient gradient boosting decision tree. Adv. Neural. Inf. Process. Syst. 30, (2017)
41. L. Prokhorenkova, G. Gusev, A. Vorobev, A.V. Dorogush, A. Gulin, CatBoost: unbiased boosting with categorical features. Adv. Neural. Inf. Process. Syst. 31, (2018)
42. P. Lerman, Fitting segmented regression models by grid search. J. R. Stat. Soc.: Ser. C: Appl. Stat. **29**(1), 77–84 (1980)
43. L. Breiman, Bagging predictors. Mach. Learn. **24**(2), 123–140 (1996)
44. F.K. Gustafsson, M. Danelljan, T.B. Schon, Evaluating scalable Bayesian deep learning methods for robust computer vision, in *Proceedings of the IEEE/CVF Conference on Computer Vision and Pattern Recognition Workshops*. (2020), pp.318–319
45. S.M. Lundberg, S.-I. Lee, A unified approach to interpreting model predictions. Adv. Neural. Inf. Process. Syst. 30, (2017)

## Publisher's Note